\documentclass[fleqn,usenatbib]{hclass}
\usepackage{newtxtext,newtxmath}

\usepackage[T1]{fontenc}
\usepackage{ae,aecompl}
\usepackage{graphicx}	
\usepackage{amsmath}
\usepackage{amssymb}	
	
\newcommand{\sty}{\scriptstyle}
\newcommand{\ssty}{\scriptscriptstyle}
\newcommand{\tsty}{\textstyle}
\newcommand{\be}{\begin{equation}}
\newcommand{\ee}{\end{equation}}
\newcommand{\lb}[1]{\label{#1}}
\newcommand{\dl}{d_{\ssty L}}
\newcommand{\da}{d_{\ssty A}}
\newcommand{\dg}{d_{\ssty G}}
\newcommand{\dz}{d_z}
\newcommand{\gaml}{\gamma_{\ssty L}}

\newcommand{\gamg}{\gamma_{\ssty G}}
\newcommand{\gamz}{\gamma_{\ssty z}}
\newcommand{\df}{\mathrm{d}}
\newcommand{\dd}{\mathrm{d}}

\newcommand{\dobs}{d_{{\tsty {\ssty \rm obs}}}}
\newcommand{\Nobs}{N_{{\tsty {\ssty \rm obs}}}}
\newcommand{\Vobs}{V_{{\tsty {\ssty \rm obs}}}}
\newcommand{\gobs}{\gamma_{{\tsty {\ssty \rm obs}}}}

\title[Fractal Analysis of the UltraVISTA Survey]{Fractal Analysis of the
UltraVISTA Galaxy Survey}
\author[S.\ Teles, A.R.\ Lopes \& M.B.\ Ribeiro]
{
Sharon Teles,$^{1}$\thanks{E-mail: steles.ts@gmail.com}
Amanda R.\ Lopes,$^{2}$$^{\mbox{\thanks{E-mail: amandalopes1920@gmail.com}}}$
and Marcelo B.\ Ribeiro$^{1,3}$$^{\mbox{\thanks{E-mail:
mbr@if.ufrj.br}\thanks{Corresponding author}}}$
\\
$^{1}$Valongo Observatory, Universidade Federal do Rio de Janeiro, Brazil\\
$^{2}$Department of Astronomy, Observat\'{o}rio Nacional, Rio de Janeiro,
      Brazil\\
$^{3}$Physics Institute, Universidade Federal do Rio de Janeiro, Brazil
}
\pubyear{2020}
\begin{document}
\label{firstpage}
\pagerange{\pageref{firstpage}--\pageref{lastpage}}
\maketitle
\begin{abstract}
This paper seeks to test if the large-scale galaxy distribution
can be characterized as a fractal system. Tools appropriate for
describing galaxy fractal structures with a single fractal
dimension $D$ in relativistic settings are developed and applied
to the UltraVISTA galaxy survey. A graph of volume-limited
samples corresponding to the redshift limits in each redshift bins
for absolute magnitude is presented. Fractal analysis using the
standard $\Lambda$CDM cosmological model is applied to a reduced
subsample in the range $0.1\le z \le 4$, and the entire sample
within $0.1\le z\le 6$. Three relativistic distances are used, the
luminosity distance $\dl$, redshift distance $\dz$ and galaxy area
distance $\dg$, because for data at $z\gtrsim 0.3$ relativistic
effects are such that for the same $z$ these distance definitions
yield different values. The results show two consecutive and
distinct redshift ranges in both the reduced and complete samples
where the data behave as a single fractal galaxy structure. For the
reduced subsample we found that the fractal dimension is $D=\left(
1.58\pm0.20\right)$ for $z<1$, and $D=\left(0.59\pm0.28\right)$ for
$1\le z\le 4$. The complete sample yielded $D=\left(1.63\pm0.20\right)$
for $z<1$ and $D=\left(0.52\pm0.29\right)$ for $1\le z\le6$. These
results are consistent with those found by \citet{gabriela}, where
a similar analysis was applied to a much more limited survey at
equivalent redshift depths, and suggest that either there are yet
unclear observational biases causing such decrease in the fractal
dimension, or the galaxy clustering was possibly more sparse and
the universe void dominated in a not too distant past. 
\end{abstract}
\section{Introduction}

The \textit{fractal galaxy distribution hypothesis} is an approach
for the description of the large-scale structure of the Universe
which assumes that this distribution is formed by a fractal system.
This approach characterizes the system by means of its key feature,
the \textit{fractal dimension} $D$, which is basically a way of
quantifying the irregularity of the distribution \citep{mandelbrot83}.
In the context of the large-scale structure of the Universe, $D$
essentially measures galactic clustering sparsity or, complementary,
the dominance of voids. Values of $D$ smaller than 3, the topological
dimension where the fractal structure is embedded, means irregular
patterns in the structure. The smaller the values of $D$, the more
sparse, or void dominated, is the galactic clustering.

The determination of the possible fractal properties of a given
galaxy distribution with data gathered from a galactic survey is
the subject of \textit{fractal analysis}, where standard techniques
of fractal geometry are applied to a given galaxy distribution
dataset in order to calculate the fractal dimension
\citep{pietronero87}. There are two ways in which this analysis can
be performed: the single fractal approach or the multifractal one
\citep{coleman92,ribeiro98}.

Describing a fractal system by means of a \textit{single fractal
dimension} is the simplest approach, since it basically reduces
the quantification of the irregular patterns within the system by
means of a unique value for $D$. The single fractal approach is also
capable of describing more complex distributions, since a system can
exhibit different values of $D$ at different distance ranges, that is,
different scaling ranges may possess different single values for $D$.
This situation means a succession of single fractal systems at
different data ranges \citep{sylos98}.

Differently from the single fractal approach, the \textit{multifractal}
one characterizes the fractal system by several fractal dimensions in
the same scaling range, that is, a whole spectrum of dimensions whose
maximum value corresponds to the single fractal dimension the structure
would have if it were treated as a single fractal. The multifractal
approach is applied when quantities like galactic luminosity or mass
have a distribution, that is, they range between very different values.
Hence, this is a generalization of $D$ that includes such distributions
and the maximum value of the multifractal spectrum corresponds to the
single fractal dimension the system would have if the studied quantity
did not range, that is, as if it did not exhibit a multifractal
pattern \citep{gabrielli2005}.

The fractal analysis of galaxy surveys whose redshift measurements are
bigger than $z\approx0.1-0.3$ cannot be done without considering
relativistic effects. This is because $D$ is determined from plots of
volume density vs.\ distance, which means volume-limited samples, and
when one determines galaxy distances at those ranges relativistic
effects become strong enough so that one is faced not with one distance
value, but several different ones whose difference increases as $z$
increases. That happens because relativistic cosmological models do not
possess a single distance definition \citep{ellis71,ellis2007,
holanda2010} and, therefore, different distance values can be assigned
for a galaxy with an empirically determined value of $z$, and the range
where those differences become significant also depends on the chosen
cosmological model. As a result, relativistic effects cannot be
neglected when performing fractal analysis of galaxy distributions
whose redshift data are gathered beyond those redshift ranges
\citep{juracy2008}.

In addition, as a consequence of relativistic effects even spatially
homogeneous cosmological models like the standard
Friedmann-Lema\^{\i}tre-Robertson-Walker (FLRW) will present
observational inhomogeneities because observations are performed along
the past light cone and at high redshift ranges distance measures
required in the determination of volume densities will necessarily
depart the local spatially homogeneous hypersurfaces of these models
\citep{ribeiro92b, ribeiro95,ribeiro2001b}. 

\textit{Fractal cosmology}, that is, modelling the large-scale structure
of the Universe by assuming that the galaxy distribution forms a fractal
system, is an old subject previously known in the literature as
\textit{hierarchical cosmology}. Discussions regarding the possible
hierarchical structuring of the Universe go as far back as the beginning
of the 20th century \citep{charlier08,charlier22,einstein22,selety22,
amoroso29}, and attempts to theoretically describe and empirically
characterize this hierarchical galaxy structure followed suit in later
decades \citep{carpenter38,vaucouleurs60,vaucouleurs70,wertz70,wertz71,
haggerty72}. The appearance of fractal geometry in the 1980s showed that
the hierarchical galaxy structure discussed in these earlier studies has
essentially the same features of galaxy distribution models that
considered this distribution as a fractal system \citep{mandelbrot83,
pietronero87}. In fact, the basic theoretical framework of these early
hierarchical models turns out to lead to the same expressions as the
ones originated from a fractal galaxy distribution model \citep{ribeiro94,
ribeiro98}.

Newtonian cosmology was used in earlier hierarchical cosmology models
\citep{wertz70,wertz71,haggerty72}, as well as in more recent fractal
cosmology ones \citep{elcio99,elcio2004}. Relativistic cosmological
models, both spatially homogeneous and inhomogeneous, which considered
a fractal system embedded in a 4-dimensional spacetime along the
observer's past light cone, producing then relativistic fractal
cosmologies, were developed later \citep{ribeiro92a,ribeiro93,
emr2001,ribeiro2001a,ribeiro2001b,ribeiro2005}. More recently, several
authors discussed relativistic cosmological models with fractal features
either theoretically \citep{mureika2004,mureika2007,sylos2011,felipe2013,
hossie2018,sadri2018,cosmai2019,jawad2019} or by attempting to apply
fractal features to different observational scenarios using Newtonian
or relativistic models \citep{jones88,martinez90,pan2000,gaite2007,
stahl2016,raj2019,bruno2020}.

Of particular interest to this paper is the study presented by
\citet[from now on CS2015]{gabriela}, which stands on the middle ground
between the two previously mentioned types of works in the sense that
it developed both theoretical tools capable of characterizing a fractal
system at high redshift by fully considering relativistic effects,
and also performed data analysis of high redshift galaxy data in order
to actually measure the single fractal dimension of the structure at
different scale ranges having deep redshift values, that is, far from
our present spatial foliation of spacetime as described in a 3+1
formalism of general relativity. This study concluded that for
$z\lesssim1.3-1.9$ the average single fractal dimension that considered
all distance definitions used in the paper resulted in $D=1.4^{\ssty
+0.7}_{\ssty-0.6}$, whereas for redshift values higher than this
approximate threshold they obtained $D=0.5^{\ssty +1.2}_{\ssty-0.4}$.

This paper aims at applying the fractal analysis methodology used by
CS2015 to a different galaxy sample, but with two major differences.
First, the fractal analysis is applied to the UltraVISTA galaxy
survey, which has measured redshift values of about 220k galaxies,
a considerably larger galaxy sample than the FORS Deep Field dataset
of 5.5k galaxies used in CS2015. Second, samples are obtained directly
from measured redshift data instead of the indirect luminosity function
methodology employed by CS2015. A graph of absolute magnitudes in terms
of redshifts shows that the UltraVISTA galaxies scale with redshift bins,
providing then a volume-limited subsample appropriate for a fractal
analysis. Nevertheless the whole survey data was also subject to fractal
analysis for comparison purposes.

As a consequence of these different approaches, the results obtained
here are clearly improved once compared to the ones achieved by CS2015,
namely, a better defined threshold for low and high scaling ranges,
smaller uncertainties and results more in line with each other
considering all cosmological distance definitions used in both studies.
Our calculations showed that summing up all results and uncertainties
obtained with the employed distance definitions we concluded that
both the subsample and entire survey data of the UltraVISTA catalog
can be well characterized as a fractal galaxy distribution system
possessing two consecutive scaling ranges with the following single
fractal dimensions. The subsample resulted in $D=\left(1.58\pm0.20
\right)$ for $z<1$, and $D=\left(0.59\pm0.28\right)$ for $1\le z\le4$,
whereas the complete sample yielded $D=\left(1.63\pm0.20\right)$ for
$0.1<z<1$, and $D=\left(0.52\pm0.29\right)$ for $1\le z\le 6$. 

The plan of the paper is as follows. Section \ref{fractal-cos}
develops standard tools of fractal geometry necessary for modelling
the large-scale galaxy distribution as a fractal system in both 
Newtonian and relativistic frameworks, comprising review material
extensively discussed and developed elsewhere plus a few additional
remarks. This is included here for a self-contained presentation.
Section \ref{fractal-analysis} describes the observational details of
the UltraVISTA galaxy survey relevant to this work, and discusses the
data handling required for the application of fractal tools to this
specific dataset. Section \ref{results} presents the results of the
fractal analysis of the UltraVISTA galaxy distribution. Discussions
and conclusions are the subject of Section~\ref{conclusion}.

\section{Fractal cosmology}\lb{fractal-cos}

It has been known since Mandelbrot's (\citeyear{mandelbrot83})
original studies that fractal systems are characterized by power-laws.
In fact, the early hierarchic cosmological models were connected to
fractals exactly because galaxy density distribution data showed
power-law features \citep[e.g.,][]{vaucouleurs70}. Hence, the early
definitions and concepts used in the hierarchical cosmologies are the
appropriate ones to start with. As it turns out these quantities form
a set of very simple concepts and definitions able to characterize
single-fractal-dimension galaxy distributions. Moreover, they are
easily and straightforwardly adapted to a relativistic setting, albeit
with some limitations.

\subsection{Newtonian hierarchical (fractal) cosmology}

Let $\Vobs$ be the \textit{observational volume} defined by the
expression below,
\be
\Vobs=\frac{4}{3} \pi {(\dobs)}^3,
\lb{vobs}
\ee
where $\dobs$ is the \textit{observational distance}. The
\textit{observed volume density} $\gobs^\ast$  is defined as
follows,
\be
\gobs^\ast=\frac{\Nobs}{\Vobs},
\lb{gobs-ast}
\ee
where $\Nobs$ is the \textit{observed cumulative number counts} of
cosmological sources, that is, galaxies. Clearly $\gobs^\ast$ gives
the number of sources per unit of observational volume out to a
distance $\dobs$.

The \textit{key hypothesis} underlining the hierarchical (fractal)
galaxy distribution relates the cumulative number counts of
observed cosmological sources and the observational distance by a
phenomenological equation called the \textit{number-distance
relation} \citep{wertz70,pietronero87}, whose expression
yields,
\be
\Nobs=B \, {(\dobs)}^D, 
\lb{Nobs}
\ee
where $B$ is a positive constant and $D$ is the fractal dimension.
This expression forms the basic hypothesis of the
\textit{Pietronero-Wertz hierarchical (fractal) cosmology}
\citep[and references therein]{ribeiro94,ribeiro98}.

Note that since $\Nobs$ is a cumulative quantity, if beyond a
certain distance there are no longer galaxies then $\Nobs$ no
longer increases with $\dobs$. If instead objects are still detected
and counted then it continues to increase. Observational biases may
possibly affect its rate of growth, leading to an intermittent
behavior, however, $\Nobs$ must grow or remain constant, and thus
the exponent in Eq.\ (\ref{Nobs}) must be positive or zero.

One may also define a second density in this context, the \textit{
observed differential density} $\gobs$. Its expression may be written
as follows \citep{wertz70,wertz71},
\be
\gobs=\frac{1}{4 \pi {(\dobs)}^2} \frac{\df N}{\dd(\dobs)}.
\lb{gobs}
\ee 
From this definition it is clear that $\gobs$ gives the rate of growth
in number counts, or more exactly in their density, as one moves along
the observational distance $\dobs$.

Substituting Eqs.\ (\ref{vobs}) and (\ref{Nobs}) into Eqs.\
(\ref{gobs-ast}) and (\ref{gobs}) we respectively reach at two forms
of the \textit{De Vaucouleurs density power-law} \citep{pietronero87,
ribeiro94}, 
\begin{equation}
\gobs^\ast = \frac{3B}{4\pi}{(\dobs)}^{D-3},
\lb{gstar3}
\end{equation}
\begin{equation}
\gobs = \frac{DB}{4\pi}{(\dobs)}^{D-3}.
\lb{gama3}
\end{equation}
Thus, if the observed galaxy distribution behaves as a fractal system
with $D<3$, that is, if it follows the number-distance relation
(\ref{Nobs}), both observational densities above decay as power-laws.
If $D=3$ the distribution is said to be \textit{observationally
homogeneous}, as both densities become constant and distance
independent. Note that the two power-laws above allow the empirical
determination of different single fractal dimensions in two or more
scaling ranges dependent on the observational distance.

The ratio between the two forms of the De Vaucouleurs density
power-law yields,
\begin{equation}
\frac{\gobs}{\gobs^\ast}=\frac{D}{3}.
\label{directD}
\end{equation}
For an observationally homogeneous galaxy distribution this ratio
must be equal to one, whereas an irregular distribution forming a
single fractal system will have $0 \leq \left( \gobs \big. \big/
\gobs^\ast \right)<1$. 

There are two important remarks to be made regarding the densities
defined by Eqs.\ (\ref{gobs-ast}) and (\ref{gobs}). First, both
$\gobs^\ast$ and $\gobs$ are radial quantities and, therefore,
should not be understood in statistical sense because they do not
average all points against all points. Second, although both
quantities are in principle equally applicable to cosmological
objects, as pointed out by \citet[Secs.\ 4.2.1, 4.2.2]{gabriela}
$\gobs$ is unsuitable for high redshift measures because the term
$\dd\Nobs/\dd(\dobs)$ in Eq.\ (\ref{gobs}) increases, reaches a
maximum and then decreases, affecting the exponent at high values
of $z$ in such a way as producing spurious negative values for $D$
at such ranges. As noted above, negative values for $D$ are not
possible due to the very definition of the number-distance relation
(\ref{Nobs}) and, hence, $\gobs$ can only be safely used at
relatively low redshift values, that is, $z\lesssim1$ or at ranges
smaller than the observed galaxy distribution histogram of galaxy
numbers per redshift bins reaches its maximum.

So, for the reasons exposed above, from now on we shall only consider
the volume density $\gobs^\ast$ in our calculations, since this
density is not contaminated by spurious effects at $z>1$.

\subsection{Relativistic fractal cosmology}

The expressions above can be applied as such in Newtonian cosmologies,
but as far as relativistic cosmological models are concerned two
important conceptual issues must be considered which alter the
expressions above in specific ways. 

First, in relativistic cosmology observations are located along
the observer's past light cone. This means that even spatially
homogeneous cosmologies like the FLRW will \textit{not} produce
observationally constant volume densities at high redshift values
because it is theoretically impossible to expect that the observed
volume density will become constant even at moderate redshift values
in FLRW cosmologies \citep[Sec.\ 2.1]{gabriela}. The basic point here
is that observational and spatial homogeneities are different
concepts in relativistic cosmology, so it is theoretically possible
to have a cosmological-principle-obeying spatially homogeneous
cosmological model exhibiting observational inhomogeneity, as
extensively discussed elsewhere \citep{ribeiro92b,ribeiro94,ribeiro95,
ribeiro2001b,ribeiro2005,juracy2008}. 

Second, both $\gobs^\ast$ and $\gobs$ are \textit{average} densities
defined in the fractal cosmology context, and, therefore, should
\textit{not} be confused with the local density appearing on the
right hand side of Einstein equations. Moreover, densities in fractal
cosmology are defined in terms of observational distances, which means
that at high redshift $\dobs$ will have different values for each
distance definition at the same redshift value $z$. In other
words, as distance in relativistic cosmology is a concept not uniquely
defined \citep{ellis71,ellis2007,holanda2010} we need to replace
$\dobs$ for $d_i$ in the equations above, where the index will
indicate the observed distance definition chosen to be calculated
with specific redshift values. In this case the applicable distance
definitions are the \textit{redshift distance} $\dz$,
\textit{luminosity distance} $\dl$ and \textit{galaxy area distance}
$\dg$, also known as \textit{transverse comoving distance}. The last
two are connected by the Etherington reciprocity law
\citep{etherington33,ellis2007} which reads as follows,
\be
\dl=(1+z)\,\dg.
\lb{eth}
\ee
The redshift distance yields, 
\be 
\dz=\frac{c \, z}{H_0},
\lb{red}
\ee
where $c$ is the light speed and $H_0$ is the Hubble constant.
This definition of $\dz$ is, of course, only valid in the FLRW
metric. \citet{vinicius2007} and \citet{iribarrem2012a} showed
that within the FLRW cosmology the densities defined by both
$\dl$ and $\dz$ have empirical power-law properties, the same
applying for $\dg$.

Another distance measure that can be defined in this context is
the \textit{angular diameter distance} $\da$, also known simply
as \textit{area distance}, which is also connected to the
quantities above by the reciprocity law, also known in the
literature as the cosmic \textit{distance duality relation}
\citep{holanda2010,holanda2011,holanda2012,zheng2020}, which
reads as follows,
\be
\dl={(1+z)}^2\da.
\lb{eth2}
\ee
However, \textit{densities} defined with $\da$ have the odd behavior
of increasing as $z$ increases, making this distance unsuitable to 
use in the context of a fractal analysis of the galaxy distribution 
\citep{ribeiro2001b,ribeiro2005,juracy2008}.

Bearing these points in mind, the expressions above become applicable
to relativistic cosmology models once they are rewritten as below,
\be
\dobs=d_i,
\lb{dists}
\ee
\be
\Vobs=V_i=\frac{4}{3} \pi {(d_i)}^3,
\lb{vi}
\ee
\be
\Nobs=N_i=B_i \, {(d_i)}^{D_i}, 
\lb{Nobs_i}
\ee
\be
\gobs^\ast=\gamma^\ast_i =\frac{N_i}{V_i}=\frac{3B_i}{4\pi}
{(d_i)}^{D_i-3},
\lb{gobs-ast_i}
\ee
where $i=({\ssty L}$, ${\sty z}$, ${\ssty G}$) according to the
distance definition used to calculate the volume density. The
proportionally constant $B_i$ of the number-distance relation
will, therefore, be attached to each specific distance definition,
this being also true for the fractal dimension $D_i$, because
$N_i$ is counted considering the limits of each distance
definition. Hence, for a given $z$ each $d_i$ yields its
respective $V_i$, $N_i$, $B_i$ and $D_i$, which means that all
quantities become attached to a certain distance definition.

As final points, first it is important to mention that although
the fractal analysis discussed above can be performed using $\dl$
in any cosmological model, the same is not true for $\dz$ because
its definition in Eq.\ (\ref{red}) is restricted to FLRW
cosmologies. Regarding $\dg$, it has been previously shown that
calculating the volume density in the Einstein-de Sitter cosmology
using $\dg$ results in $\gamg^\ast=\mbox{constant}$ \citep[pp.\
1718, 1723, 1724]{ribeiro2001b}, which seemed to indicate $\dg$ as
an unsuitable distance definition to be used in fractal analysis.
Nevertheless, later works showed that such behavior is cosmology
dependent, not being valid in other FLRW models \citetext{\citealp{
vinicius2007}, fig.\ 7; \citealp{iribarrem2012a}, figs.\ 2-5;
\citealp{gabriela}, figs.\ 3-4}. This finding justifies the
inclusion of $\gamg^\ast$ in the present study. 

Second, the concepts above indicate that reasoning that regards the
fractal approach to the galaxy distribution only at low redshift ranges
are not applicable to the analysis performed in this paper. As the
light cone is a relativistic concept, confusion arises if one does
not acknowledge the difference between spatial and observational
volume densities. It is in the latter spacetime locus that astronomy
is made and where fractality in the sense of this work may be
detected. Hence, observed fractal features appear only by correctly
manipulating the FLRW observational quantities along the observer's
past light cone at ranges where the null cone effects become relevant
as far as distance definitions are concerned \citep{ribeiro95,
ribeiro2001b,ribeiro2005,juracy2008}. These effects form the very
core of the present analysis.

\section{Fractal analysis}\lb{fractal-analysis}

As this study seeks to empirically ascertain whether or not the
fractal galaxy distribution hypothesis holds at very large-scales
of the observed Universe, we chose to perform a fractal analysis
with the data provided by the UltraVISTA galaxy survey, since it
contains hundreds of thousands of galaxies with measured
redshifts. Let us present next some details about this survey and
how the fractal analysis was performed in their data. 

\subsection{The UltraVISTA galaxy survey}\lb{uvista}

\begin{figure}
  \includegraphics[width=\columnwidth]{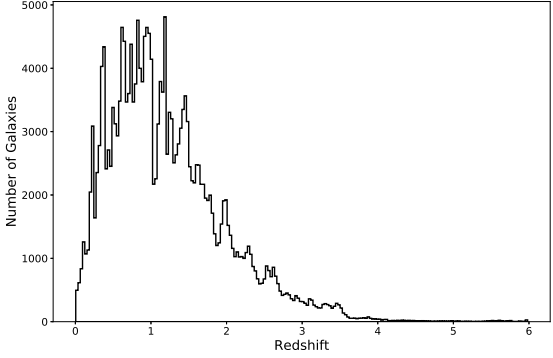}
  \caption{Histogram showing the galaxy distribution numbers in terms
  of redshift for the UltraVISTA survey dataset studied here. This
  graph has $\Delta z=0.03$ as the redshift bins' size.}
  \label{histz}
\end{figure}

\begin{figure}
  \includegraphics[width=\columnwidth]{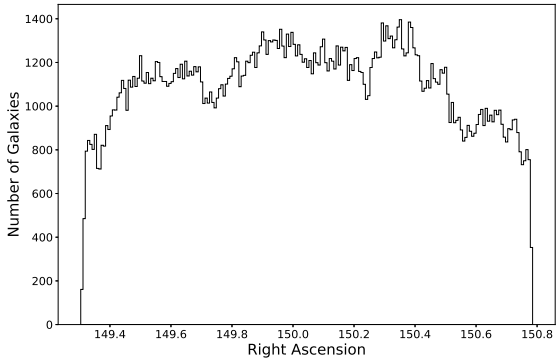}
  \caption{Histogram showing the galaxy distribution numbers in terms
  of right ascension (deg) for the UltraVISTA survey dataset studied here.}
  \label{histra}
\end{figure}

\begin{figure}
  \includegraphics[width=\columnwidth]{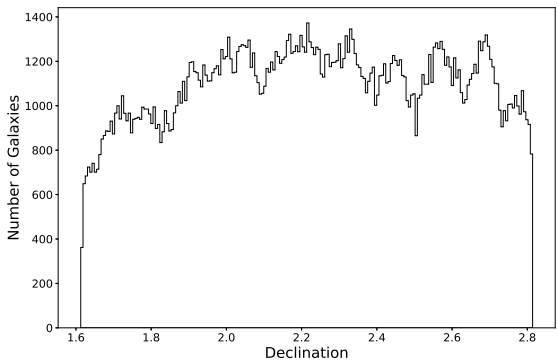}
  \caption{Histogram showing the galaxy distribution numbers in terms
  of declination (deg) for the UltraVISTA survey dataset studied here.}
  \label{histd}
\end{figure}

Our data are based on the first data release of the UltraVISTA
galaxy survey, which is centered on the COSMOS field
\citep{cosmos2007} with an effective area of $1.5\,\mbox{deg}^{\,2}$.
Observations were made in 4 near infrared filters, $Y$, $J$, $H$
and $K_{\mathrm{S}}$, described in \citet{ultra1}. Photometric
redshifts were calculated by \citet{ultra11} applying the SED
fitting technique to 29 bands, the ones from UltraVISTA and a
complementary set of broad and narrow bands from other surveys
encompassing the ultraviolet, optical, near infrared and mid
infrared regimes. The initial dataset consisted in a
$K_{\mathrm{S}}$-band selected ($K_{\mathrm{S}}<24$) sample of
about 220k galaxies in the redshift range of $0.1\le z\le6$.
Although the sample was originally divided into quiescent and
star forming galaxies, such grouping is unimportant for the
purposes of this study and, hence, all galaxies of both types were
included in the subsample selection and analysis performed here.
Fig.\ \ref{histz} shows the UltraVISTA survey's galaxy numbers
distribution in terms of redshift. Figs.\ \ref{histra} and
\ref{histd} respectively show the galaxy numbers distribution in
terms of right ascension and declination.

\subsection{Data selection}\lb{dataselec}

The fractal analytical tools discussed above require the establishment
of volume limited samples. However, galaxy surveys are limited by
apparent magnitude, so some methodology is needed to ascertain that
reduced subsamples follow increasing redshift bins in order to
render the reduced galaxy data distributed over limited volume bins.
One way of doing this is by plotting the galaxies' absolute magnitudes
in terms of their respective measured redshifts and then selecting
galaxies below a certain absolute magnitude threshold defined by the
limiting apparent magnitude of the survey. This can be done by using
the usual expression below,
\be
M=m-5\log \dl (z) -25,
\lb{magabs1}
\ee
where $M$ is the absolute magnitude, $m$ is the apparent magnitude
and $\dl$ is given in Mpc. Next one needs to choose the apparent
magnitude threshold $m$ and its passband, as well as verifying if
the resulting data is indeed, at least generally, distributed along
the measured redshift bins, and then possibly establishing a redshift
window for the final subsample distribution.

\begin{figure}
  \includegraphics[width=\columnwidth]{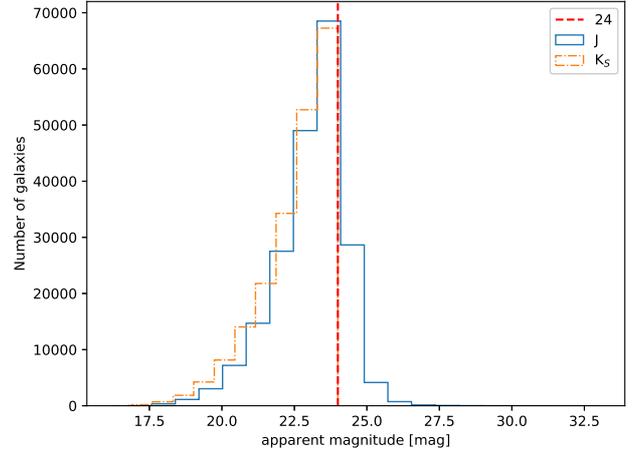}
  \caption{UltraVISTA galaxy numbers plotted in terms of apparent 
           magnitudes in both the $J$ and $K_{\mathrm{S}}$ passbands.
           The number distribution for both bandwidths peaks at
           apparent magnitudes equal to 24 indicated by the vertical
           line.}
  \label{hist-number-mags}
\end{figure}

The UltraVISTA survey furnished absolute magnitudes calculated in
the $NUV$, $B$, $R$ and $J$ passbands. The $J$-band is the best
choice among those for our purposes here, since this near infrared
filter is less affected by dust extinction. Fig.\
\ref{hist-number-mags} shows an histogram of the UltraVISTA galaxy
numbers in terms of apparent magnitudes in the $K_{\mathrm{S}}$ and
$J$ passbands. Clearly the number distribution peaks at the apparent
magnitude value 24 for both wavebands, a fact that led us to choose
$J=24$ as our apparent magnitude threshold. Therefore, Eq.\
(\ref{magabs1}) can be rewritten according to the expression below,
which provides the dividing line between the selected and unselected
galaxies.
\be
M_J=24-5\log \dl (z) -25.
\lb{magabs2}
\ee

\begin{figure}
  \includegraphics[width=\columnwidth]{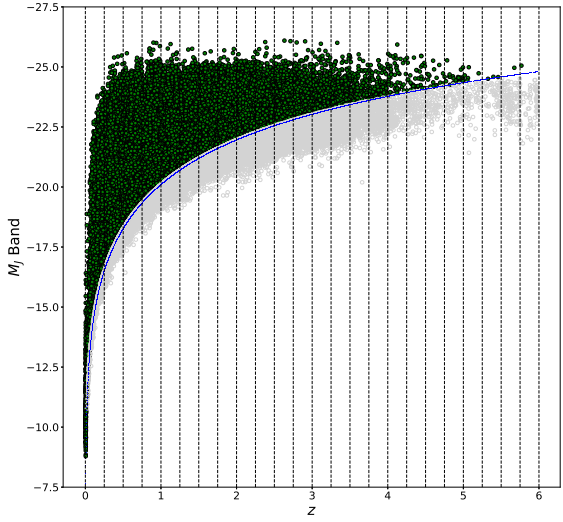}
  \caption{Absolute magnitudes in the $J$-band vs.\ redshift for all
           galaxies of the UltraVISTA survey. Green filled
           circles are galaxies having $M_J$ smaller than the
           blue line cut given by Eq.\ (\ref{magabs2}), whereas
           open gray circles have bigger values for $M_J$.}
  \label{mvsz}
\end{figure}

Fig.\ \ref{mvsz} shows a plot of the absolute magnitudes of 219300
UltraVISTA galaxies in the $J$-band against their respective
measured redshifts. Green filled circles are those whose absolute
magnitudes provided by the survey are smaller than $M_J$ in Eq.\
(\ref{magabs2}), and are then inside the reduced subsample,
whereas the open gray circles are outside this threshold. We
noticed that the absolute magnitude cut based on the $J$-band
corresponds to a volume-limited subsample because the data
generally follow the redshift bins of increasing values. In
addition, we also noticed that there are few galaxies at the
tail of the distribution, so our subsample suffered a further
cut at $z=4$.

Hence, we end up with a \textit{reduced UltraVISTA subsample} of
166566 galaxies cut by $J$-band absolute magnitudes $M_J$ and
limited up to $z=4$. The remaining 52734 galaxies outside this
subsample, that is, below the dividing line in Fig.\ \ref{mvsz}
and having $z>4$, were disregarded.

\subsection{Data analysis}\lb{data-analysis}

To obtain the observational distances $d_i\,(i={\ssty G}$,
${\ssty L}$, ${\sty z})$ from the calculated photometric redshift 
values one needs to choose a cosmological model. We assumed the
FLRW cosmology with $\Omega_{m_{\ssty 0}}=0.3$,
$\Omega_{\Lambda_{\ssty 0}} = 0.7$ and $H_0=70 \; \mbox{km} \;
{\mbox{s}}^{-1} \; {\mbox{Mpc}}^{-1}$.

The next steps were the establishment of the minimum redshift value
$z_{\ssty 0}$ from where to start the analysis, the respective
minimum distances $d_{i_0}=d_{i_0}(z_{\ssty 0})$, and the incremental
distance interval $\Delta d_i$. The data sorting process was
initiated by counting the number of observed galaxies
$N_{i_{\ssty \rm 1}}$ in the first interval $d_{i_1}=d_{i_0}+\Delta
d_i$ and calculating the respective volume density $\gamma_{i_1}^\ast$.
This first interval defined the first bin. Then, the size of the bin
was increased by $\Delta d_i$ and values for $N_{i_{\ssty \rm 2}}$
and $\gamma_{i_2}^\ast$ were obtained at the distance
$d_{i_2}=d_{i_0}+2\Delta d_i$. This algorithm was repeated $n$
times until the last, and farthest, group of galaxies were included
and all relevant quantities were also counted and calculated.

Different bin size increments $\Delta d_i$ were tested for each
distance definition in order to find out whether or not that would
affect the results. This test turned out negative, which means that
the obtained results are independent of bin size increment. We then 
chose $\Delta d_i=200$~Mpc, value which was applied to all
calculations and provided in the end a very reasonable amount of
data points for each quantity from where simple linear regression
analyses were able to be performed.

The final step was the determination of the fractal dimension
itself. If the galaxy distribution really formed a fractal system,
according to Eq.\ (\ref{gobs-ast_i}) the graphs of $\gamma_i^\ast$
vs.\ $d_i$ would behave as decaying power-law curves, and whose
linear fit slopes in log-log plots allow for the fractal dimensions
$D_i$ of the distribution to be straightforwardly determined.

\section{Results}\lb{results}

\subsection{Reduced subsample}

Graphs for log-log values of $\gamma_i^\ast$ vs.\ $d_i$ showed
that to a good approximation the reduced UltraVISTA galaxy
survey subsample sorted according to the criteria set at Sec.\
\ref{dataselec} conforms with what is predicted as if the
galaxy distribution does form a fractal system. Moreover, two
power-law decaying regions were observed in the data, for $z<1$
and $z>1$, producing different single fractal dimensions.

\begin{figure}
  \includegraphics[width=\columnwidth]{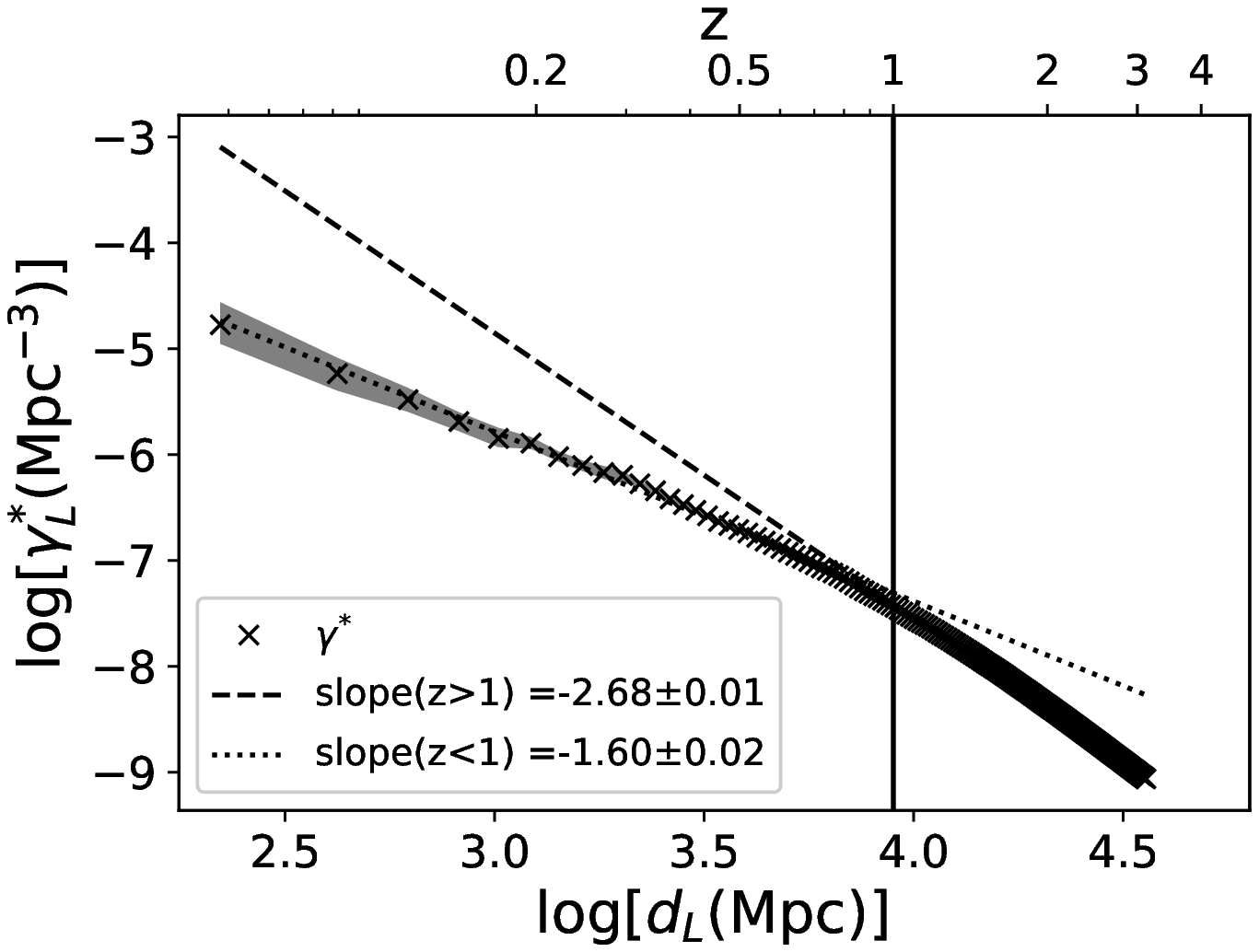}
  \caption{Graph showing the log-log results for $\gaml^\ast$
	  vs.\ $\dl$ obtained with the \textit{reduced}
	  UltraVISTA galaxy redshift survey dataset (see Sec.\
	  \ref{dataselec}). The dotted line is the straight
	  line fit for galaxies having $z<1$, whereas the
	  dashed line is for those with $z>1$. The error
	  area is in gray. According to Eq.\ (\ref{gobs-ast_i})
	  the fractal dimensions obtained from these data are
	  $D_{\ssty L}=(1.40\pm0.02)$ for $z<1$ and $D_{\ssty
          L}=(0.32 \pm0.01)$ for $1\le z\le 4$.} 
  \label{gammaLdL-r}
\end{figure}

\begin{figure}
  \includegraphics[width=\columnwidth]{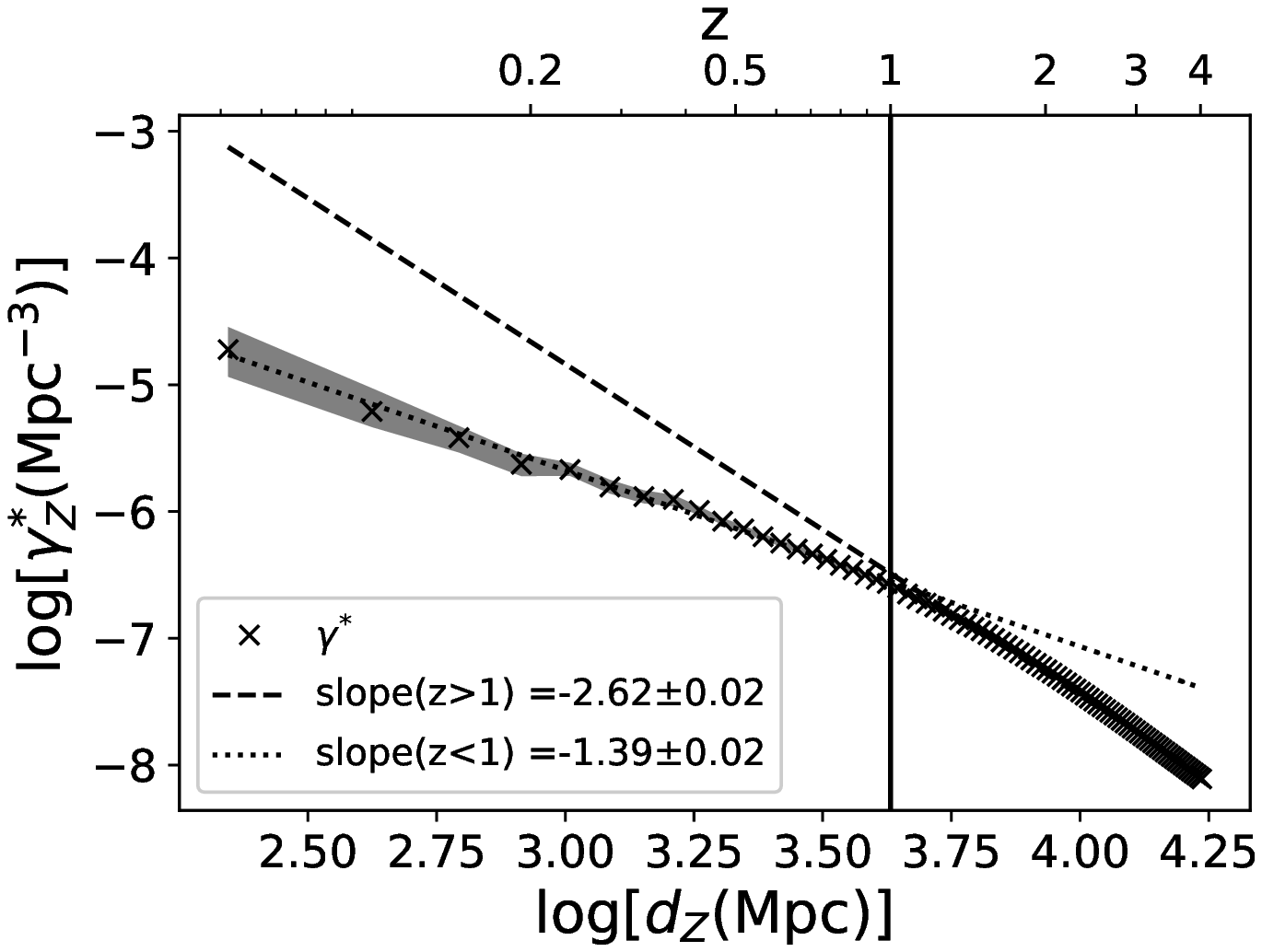}
  \caption{Graph showing the log-log results for $\gamz^\ast$
	  vs.\ $\dz$ obtained with the \textit{reduced}
	  UltraVISTA galaxy redshift survey dataset (see Sec.\
	  \ref{dataselec}). The dotted line is the straight
	  line fit for galaxies having $z<1$, whereas the
	  dashed line is for those with $z>1$. The error
	  area is in gray. According to Eq.\ (\ref{gobs-ast_i})
	  the fractal dimensions obtained from these data are
	  $D_{\ssty z}=(1.61\pm0.02)$ for $z<1$ and $D_{\ssty
          z}=(0.38 \pm0.02)$ for $1\le z\le 4$.} 
  \label{gammaZdZ-r}
\end{figure}

\begin{figure}
  \includegraphics[width=\columnwidth]{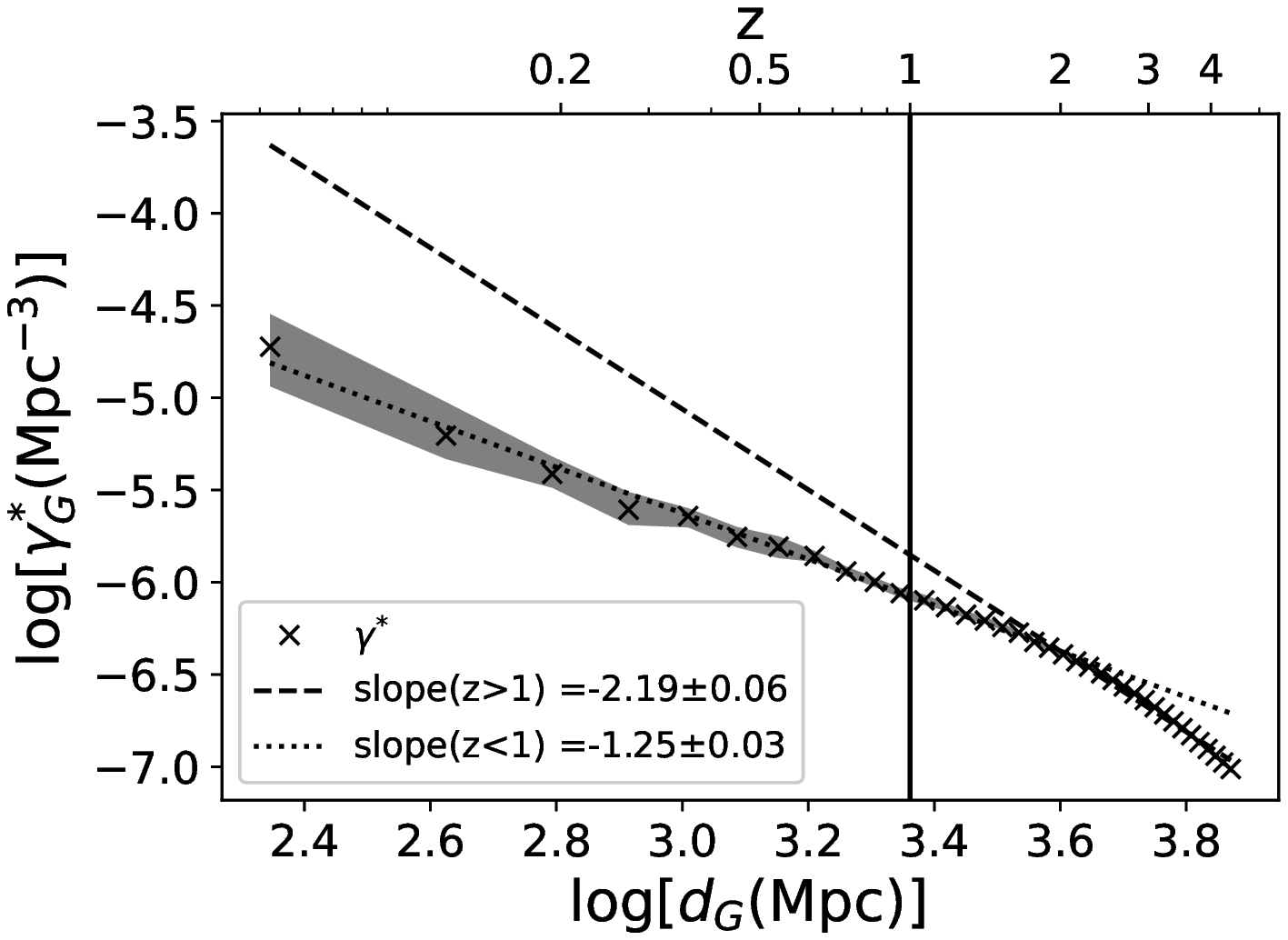}
  \caption{Graph showing the log-log results for $\gamg^\ast$
	  vs.\ $\dg$ obtained with the \textit{reduced}
	  UltraVISTA galaxy redshift survey dataset (see Sec.\
	  \ref{dataselec}). The dotted line is the straight
	  line fit for galaxies having $z<1$, whereas the
	  dashed line is for those with $z>1$. The error
	  area is in gray. According to Eq.\ (\ref{gobs-ast_i})
	  the fractal dimensions obtained from these data are
	  $D_{\ssty G}=(1.75\pm0.03)$ for $z<1$ and $D_{\ssty
          G}=(0.81 \pm0.06)$ for $1\le z\le 4$.} 
  \label{gammaGdG-r}
\end{figure}

Figs.\ \ref{gammaLdL-r} to \ref{gammaGdG-r} present the results
for all distance definitions adopted here, from where it can be
concluded that for $z<1$ the fractal dimension is in the range
$1.38-1.78$, whereas for $1\le z\le4$ the resulting range is
significantly smaller, $0.31-0.87$. Table \ref{tab1} collects
all results.
\begin{table}
\caption{Results in two redshift scales of the UltraVISTA
	galaxy survey fractal analysis in the \textit{reduced}
	subsample (see Sec.\ \ref{dataselec}). The single fractal
	dimensions $D_{\ssty L}$, $D_{\ssty z}$ and $D_{\ssty
	G}$ were obtained from this galaxy distribution
	respectively using the luminosity distance $\dl$,
	redshift distance $\dz$ and galaxy area distance
        (transverse comoving distance) $\dg$.}
\label{tab1}
\begin{center}
\begin{tabular}{cccc}
\hline
& $D_{\ssty L}$ & $D_{\ssty z}$ & $D_{\ssty G}$\\
\hline
$z<1$         & $1.40\pm0.02$ & $1.61\pm0.02$ & $1.75\pm0.03$\\
$1\le z\le 4$ & $0.32\pm0.01$ & $0.38\pm0.02$ & $0.81\pm0.06$\\
\hline
\end{tabular}
\end{center}
\end{table}

In summary, the results above indicate that the UltraVISTA galaxy
survey provided a galaxy distribution subsample dataset that can
be described as a fractal system with the following two consecutive
and approximate single fractal dimensions values: $D\,(z<1)=(1.58
\pm0.20)$ and $D\,(1\le z\le4)=(0.59\pm0.28)$. The possible reasons
as why the fractal dimension is so much reduced at the deepest range
$z>1$ will be discussed below.

\subsection{Complete (unselected) survey data}

The fractal analysis of the UltraVISTA galaxy subsample, as
defined in Sec.\ \ref{dataselec}, were based in the plot of
absolute magnitudes in the $J$-band in terms of the measured
redshifts of the galaxies shown in Fig.\ \ref{mvsz}. However,
this plot also shows that even the galaxies outside the absolute
magnitude cut are also distributed along increasing redshift bins,
a fact that suggests that the whole sample may also be
volume-limited. Hence, it is interesting to apply the same fractal
methodology developed above not only to the subsample, but also to
the whole survey data in order to compare the results.

\begin{figure}
  \includegraphics[width=\columnwidth]{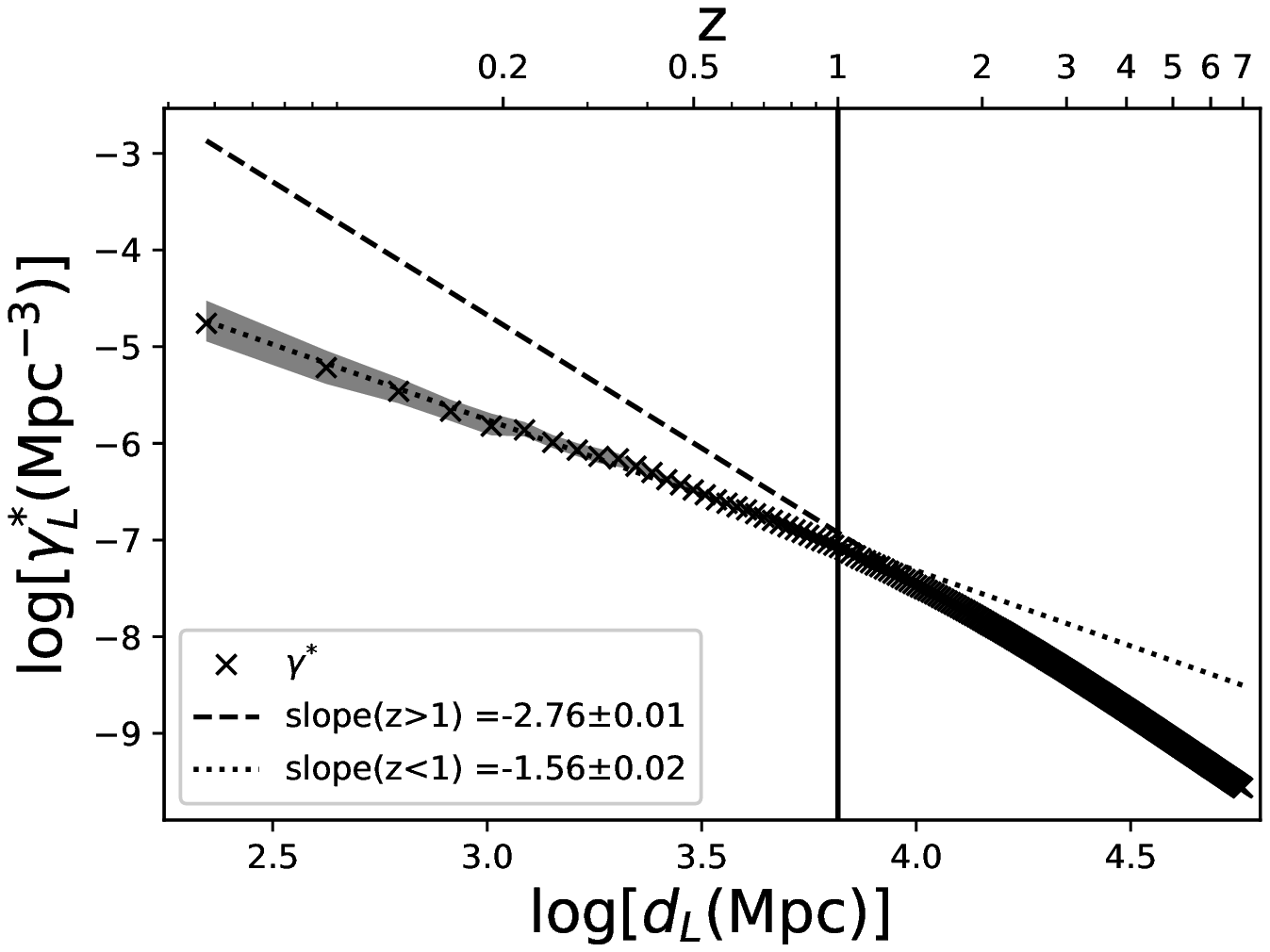}
  \caption{Graph showing the log-log results for $\gaml^\ast$
	  vs.\ $\dl$ obtained with the complete UltraVISTA galaxy
	  survey dataset. The dotted line is the
	  straight line fit for galaxies having $z<1$, whereas
	  the dashed line is for those with $z>1$. The error
	  area is in gray. According to Eq.\ (\ref{gobs-ast_i})
	  the fractal dimensions obtained from these data are
	  $D_{\ssty L}=(1.44\pm0.02)$ for $z<1$ and $D_{\ssty
          L}=(0.24 \pm0.01)$ for $1\le z\le 6$.} 
  \label{gammaLdL}
\end{figure}

\begin{figure}
  \includegraphics[width=\columnwidth]{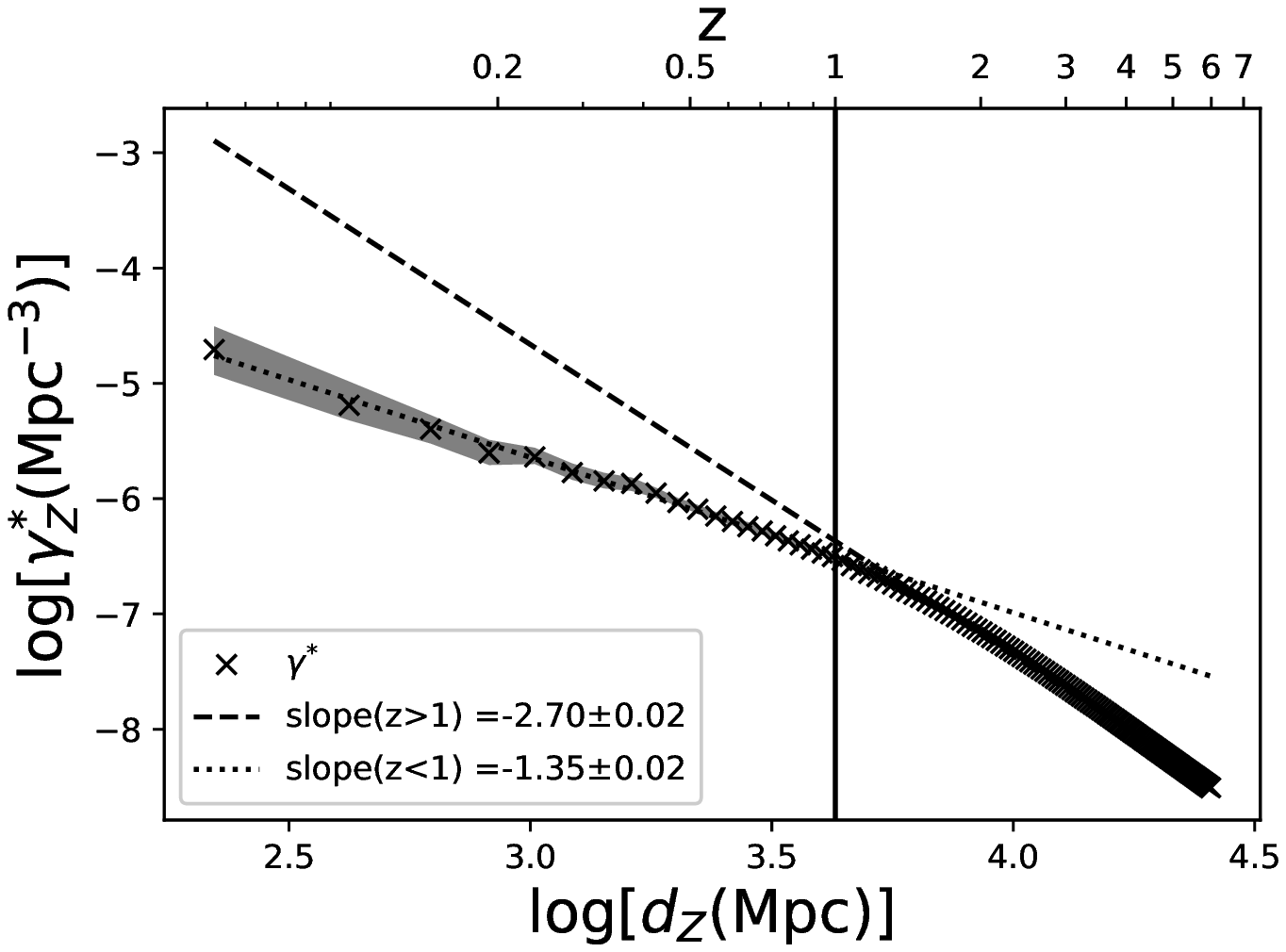}
  \caption{Graph showing the log-log results for $\gamz^\ast$
	  vs.\ $\dz$ obtained with the complete UltraVISTA galaxy
	  survey dataset. The dotted line is the
	  straight line fit for galaxies having $z<1$, whereas
	  the dashed line is for those with $z>1$. The error
	  area is in gray. According to Eq.\ (\ref{gobs-ast_i})
	  the fractal dimensions obtained from these data are
	  $D_{\ssty z}=(1.65\pm0.02)$ for $z<1$ and $D_{\ssty
          z}=(0.30 \pm0.02)$ for $1\le z\le 6$.} 
  \label{gammaZdZ}
\end{figure}

\begin{figure}
  \includegraphics[width=\columnwidth]{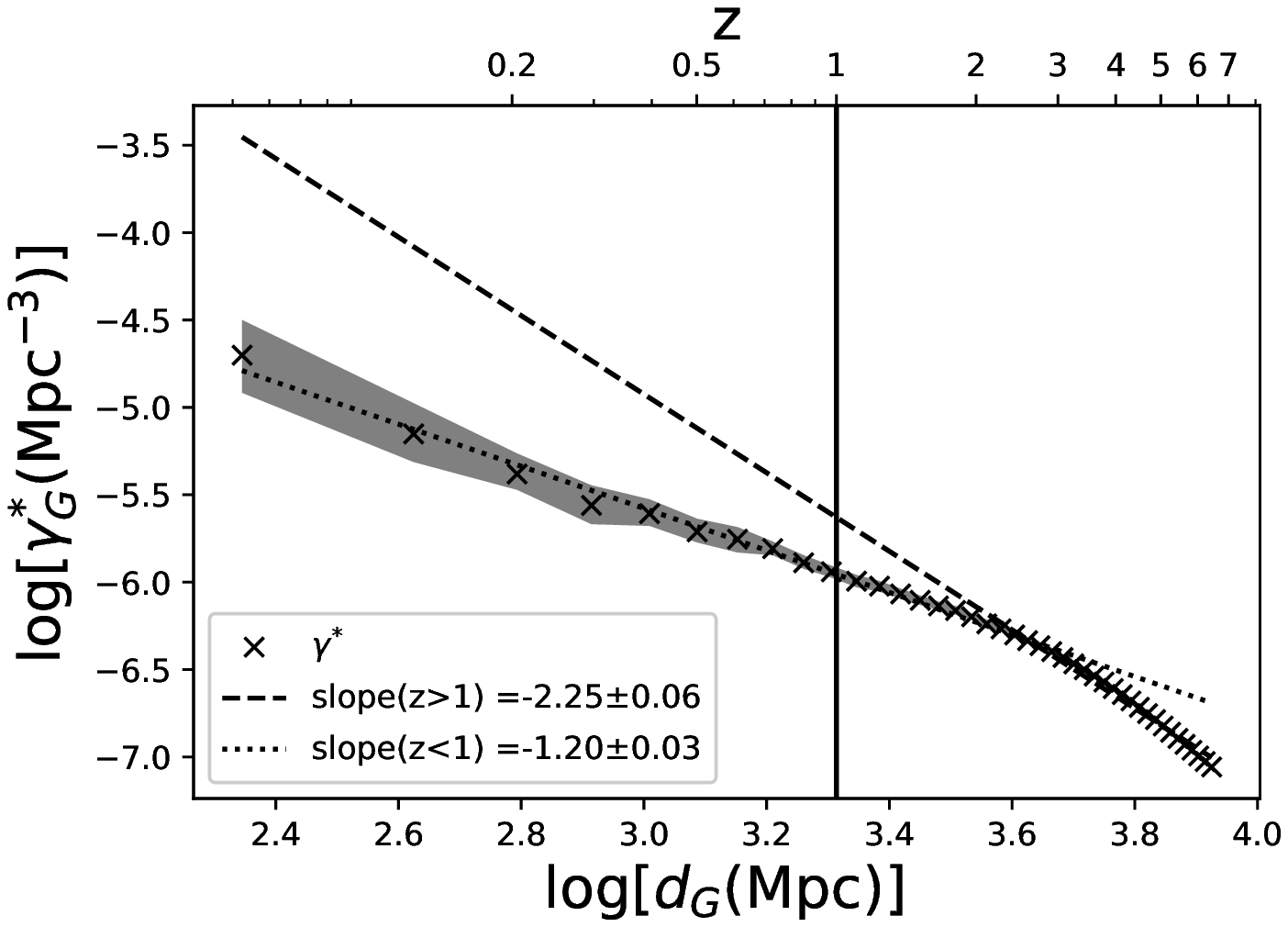}
  \caption{Graph showing the log-log results for $\gamg^\ast$
	  vs.\ $\dg$ obtained with the complete UltraVISTA galaxy
	  survey dataset. The dotted line is the
	  straight line fit for galaxies having $z<1$, whereas
	  the dashed line is for those with $z>1$. The error
	  area is in gray. According to Eq.\ (\ref{gobs-ast_i})
	  the fractal dimensions obtained from these data are
	  $D_{\ssty G}=(1.80\pm0.03)$ for $z<1$ and $D_{\ssty
          G}=(0.75 \pm0.06)$ for $1\le z\le 6$.} 
  \label{gammaGdG}
\end{figure}

Figs.\ \ref{gammaLdL}-\ref{gammaGdG} show the results of the
fractal analysis of the complete UltraVISTA survey data. It
is clear that the galaxy distribution also presents fractal
features in two regions below and above the redshift value $z=1$.
The corresponding single fractal dimensions for each distance
definitions adopted here were found to lie in the range $1.42-1.83$
for $z<1$, whereas for $1\le z\le6$ the dimension is significantly
smaller, in the range $0.23-0.81$. Table \ref{tab2} collects these
results and a comparison with the ones for the reduced subsample
presented in Table \ref{tab1} indicated results very much alike,
although their respective uncertainties do not overlap.

So, it seems that the whole UltraVISTA galaxy survey can also
be described as a fractal system having two consecutive single
fractal dimensions: $D\,(z<1)=(1.63\pm0.20)$ and $D\,(1\le z\le6)
=(0.52\pm0.29)$. 
\begin{table}
\caption{Results in two redshift scales of the UltraVISTA
	galaxy survey fractal analysis in the \textit{complete}
	sample. The single fractal dimensions $D_{\ssty L}$,
	$D_{\ssty z}$ and $D_{\ssty G}$ were obtained from this
	galaxy distribution respectively using the luminosity
	distance $\dl$, redshift distance $\dz$ and galaxy area
	distance (transverse comoving distance) $\dg$.}
\label{tab2}
\begin{center}
\begin{tabular}{cccc}
\hline
& $D_{\ssty L}$ & $D_{\ssty z}$ & $D_{\ssty G}$\\
\hline
$z<1$         & $1.44\pm0.02$ & $1.65\pm0.02$ & $1.80\pm0.03$\\
$1\le z\le 6$ & $0.24\pm0.01$ & $0.30\pm0.02$ & $0.75\pm0.06$\\
\hline
\end{tabular}
\end{center}
\end{table}

\section{Conclusions}\lb{conclusion}

This paper sought to empirically test if the large-scale galaxy
distribution can be described as a fractal system. Tools originally
developed for Newtonian hierarchical cosmology were extended and
applied to relativistic cosmological models in order to possibly
describe galaxy fractal structures by means of single fractal
dimensions at deep redshift values. These tools were applied to the
UltraVISTA galaxy survey dataset comprising 220k objects spanning
the redshift interval of $0.1\le z\le 6$.

A reduced subsample of the survey was established by plotting the
galaxies' absolute magnitudes in the $J$-band against their
respective redshifts and setting an absolute magnitude cut with the
apparent magnitude of $J=24$, as well as a redshift cut at $z=4$. 
Since this subsample showed that its galaxies followed increasing
redshift bins, they were considered as effectively being a
volume-limited distribution.

Fractal analysis of the reduced subsample was carried out using the
standard $\Lambda$CDM relativistic cosmological model. As
relativistic cosmologies have several definitions of observed
distance, only three distinct ones were used here, namely the
luminosity distance $\dl$, redshift distance $\dz$ and galaxy area
distance $\dg$, also known as transverse comoving distance. The use
of several cosmological distance measures is due to the fact that
relativistic effects become strong enough for redshift ranges
larger than $z\gtrsim 0.3$, so these distance definitions produce
different results for the same redshift value at those ranges. An
algorithm for sorting the data according to the analytical
developments required for testing an observed fractal structure as
discussed here was also detailed.

The results indicate that the UltraVISTA subsample has two
consecutive redshift ranges behaving as single fractal structures.
For $z<1$ the derived fractal dimension is well approximated by
$D=(1.58\pm0.20)$, whereas for $1\le z\le4$ this dimension
decreased to $D=(0.59\pm0.28)$. For comparison, the
same fractal analysis was also carried out in the complete survey
data, yielding as well two consecutive redshift ranges
characterized as single fractal systems: $D=(1.63\pm0.20)$ for
$z<1$ and $D=(0.52\pm0.29)$ in the range $1\le z\le6$. Both results,
from the reduced and complete data, are consistent with those found
by \citet{gabriela}, although here the conclusions were reached by
a different, and simpler, methodology, and by applying this
methodology to a numerically much larger galaxy sample in both
cases.

The most obvious question regarding these results is why there is
such a significant decrease in the fractal dimension for redshift
values larger than the unity. Conceivably this might be due to an
observational bias caused by the simple fact that many galaxies
located beyond $z=1$ are not being detected, decreasing then the
observed galaxy clustering and, therefore, the associated fractal
dimension at those scales. One might also consider the possibility
of a bias in the galaxy number counts due to the small angular area
of this survey. This means that its observational area might not
provide a representative measurement of the entire sky distribution.
Moreover it is important to consider the choice of observational
field. If the size of the observed area is small and contains galaxy
clusters, some galaxy concentration peaks can lead to higher fractal
dimensions at low $z$.

It is also conceivable that the data used to obtain our results suffer
from a detection bias related to the cumulative measure $\Nobs$.
Unbiassing this quantity might depend on comparing observed $\Nobs$ data
with simulations derived from cosmological models' subsets and halo
occupation distribution models \citep{hod}. One must also remember that
the UltraVISTA photometric redshifts were obtained using SED fitting
analysis, which means that the errors in the values are entirely
dependent on the input assumptions used to derive these photo-$z$, and,
therefore, it is not clear how the use of different types of
photometric modeling might affect our results. This possible error
source may be mitigated by the use of restricted galaxy samples, say,
containing only luminous red galaxies such that photometric redshift
uncertainties are minimized \citep{redmagic,redmagic2}.

Another possible source of data uncertainty that might affect
our results is the tension among the several measures of the Hubble
constant. Here we have assumed the standard FLRW cosmological model
with $H_0=70\;\mbox{km}\;{\mbox{s}}^{-1}{\mbox{Mpc}}^{-1}$,
nevertheless different Hubble constant predictions range this quantity
by about $\pm4\;\mbox{km}\;{\mbox{s}}^{-1}{\mbox{Mpc}}^{-1}$.
Relativistic cosmological models are known to be highly nonlinear, so
how much this uncertainty in the Hubble constant might affect our
results, if at all, cannot be quantified beforehand. So, to ascertain
the possibly negligible, or otherwise, impact of this uncertainty in
our results can only be concluded by actually carrying out the
calculations with this parameter range and performing comparisons.

Aside from possible observational biases and yet unknown impact due
to error sources, one might also attribute the decrease in the
fractal dimension to a real physical effect. Perhaps, galaxy evolution
dynamics is at play in causing such a decrease in the sense that
there might be indeed much less galaxies at high $z$, meaning that
the Universe was void dominated at those epochs since galaxies were
much more sparsely distributed and in smaller numbers. Only further
work with different high-$z$ galaxy samples and in different regions
of the sky may clarify this issue.

\section*{Acknowledgments}

We are grateful to the referee for useful comments. S.T.\ thanks
the Universidade Federal do Rio de Janeiro for a PIBIC scholarship.
A.R.L.\ acknowledges Brazil's Federal Funding Agency CNPq for the
financial support with a PCI fellowship.

\bibliographystyle{h}
\bibliography{cosmo} 
\bsp	
\label{lastpage}
\end{document}